# Network reinforcement driven drug repurposing for COVID-19 by exploiting disease-gene-drug associations

Yonghyun Nam[1], Jae-Seung Yun[1,2], Seung Mi Lee[1,3], Ji Won Park[1,4], Ziqi Chen[5], Brian Lee[1], Anurag Verma[6], Xia Ning[5,7,8], Li Shen[1,9], Dokyoon Kim[1,9]*

[1]Department of Biostatistics, Epidemiology and Informatics, Perelman School of Medicine, University of Pennsylvania, Philadelphia, USA

[2]Division of Endocrinology and Metabolism, Department of Internal Medicine, St. Vincent's Hospital, College of Medicine, The Catholic University of Korea, Seoul, South Korea

[3]Department of Obstetrics and Gynecology, Seoul National University College of Medicine, Seoul, South Korea

[4]Department of Surgery, Seoul National University College of Medicine, Seoul, South Korea

[5]Computer Science and Engineering Department, College of Engineering, The Ohio State University, Columbus, USA

[6]Department of Genetics, Perelman School of Medicine, University of Pennsylvania, Philadelphia, USA

[7]Biomedical Informatics Department, College of Medicine, The Ohio State University, Columbus, USA

[8] Translational Data Analytics Institute, The Ohio State University, Columbus, USA

[9]Institute for Biomedical Informatics, University of Pennsylvania, Philadelphia, USA

*corresponding author

**Abstract**

Currently, the number of patients with COVID-19 has significantly increased. Thus, there is an urgent need for developing treatments for COVID-19. Drug repurposing, which is the process of reusing already-approved drugs for new medical conditions, can be a good way to solve this problem quickly and broadly. Many clinical trials for COVID-19 patients using treatments for other diseases have already been in place or will be performed at clinical sites in the near future. Additionally, patients with comorbidities such as diabetes mellitus, obesity, liver cirrhosis, kidney diseases, hypertension, and asthma are at higher risk for severe illness from COVID-19. Thus, the relationship of comorbidity disease with COVID-19 may help to find repurposable drugs. To reduce trial and error in finding treatments for COVID-19, we propose building a network-based drug repurposing framework to prioritize repurposable drugs. First, we utilized knowledge of COVID-19 to construct a disease-gene-drug network (DGDr-Net) representing a COVID-19-centric interactome with components for diseases, genes, and drugs. DGDr-Net consisted of 592 diseases, 26,681 human genes and 2,173 drugs, and medical information for 18 common comorbidities. The DGDr-Net recommended candidate repurposable drugs for COVID-19 through network reinforcement driven scoring algorithms. The scoring algorithms determined the priority of recommendations by utilizing graph-based semi-supervised learning. From the predicted scores, we recommended 30 drugs, including dexamethasone, resveratrol, methotrexate, indomethacin, quercetin, etc., as repurposable drugs for COVID-19, and the results were verified with drugs that have been under clinical trials. The list of drugs via a data-driven computational approach could help reduce trial-and-error in finding treatment for COVID-19.

## 1. Introduction

Coronavirus typically affects the respiratory tract of humans, and it leads to mild severe respiratory tract infections. Since December 2019, the novel coronavirus, well known as severe acute respiratory syndrome coronavirus 2 (SARS-COV-2) or coronavirus disease 2019 (COVID-19), has been declared a global pandemic by the World Health Organization[1]. As of June 14th, there are about 755 million confirmed cases and about 42.3 million deaths worldwide. The most common symptoms of COVID-19 are fever, dry cough, and fatigue. Additional symptoms include nasal congestion, headaches, conjunctivitis, and sore throat. In severe cases, symptoms of dyspnea and chest pain were reported.

Several recent studies analyzed hospitalized patients' records to investigate diseases comorbid with COVID-19. Patients with these comorbidities such as diabetes mellitus, obesity, liver cirrhosis, kidney diseases, hypertension, and asthma have a higher risk for severe illness from COVID-19[2-7]. Among confirmed cases of COVID-19, patients with any of the above comorbidities do worse than those without. Since the coronavirus leads to death in severe cases and is highly contagious between humans, there is an urgent need for developing therapeutic treatments or vaccines. However, there are currently no effective medications[8].

To overcome the pandemic, many researchers and companies are currently developing new treatments and vaccines for COVID-19. However, this process usually takes at least a decade of work and about $300-600 million of funding. In order to overcome this global pandemic quickly, treatments must be developed immediately, but this can be difficult if we only employ traditional drug discovery methods. One way to hasten our search for effective treatments involves drug repurposing. Drug repurposing (or drug repositioning) is the process of reusing of already-approved drugs for new medical conditions. Many clinical trials for COVID-19 patients using treatments for other diseases are already in place or will be performed in the clinical sites shortly[10]. So far, candidate treatments such as Remdesivir (originally designed to treat the Ebola virus), hydroxychloroquine (the now-infamous malaria drug), and lopinavir/ritonavir (an antiretroviral drug targeting HIV) have been used on hospitalized COVID-19 patients. In addition to these clinical trials, many studies have recently been conducted using computational drug repurposing strategies. Zhou et al constructed a drug-human coronavirus network using protein associations between coronavirus and 2,938 drugs and discovered 16 repurposable drugs[11]. They constructed a protein interactome network using coronavirus and drug-related genes and then predicted candidate drugs using network proximities. Ge et al proposed a data-driven drug repurposing framework that integrates coronavirus-related data and identifies drug candidates from 6,255

drugs[12]. Muralidharan et al found candidate drugs capable of molecular docking or drug binding by modeling the protein structure of coronavirus and drugs with structure-based virtual screening identified four commercially available drugs[13]. These previous studies used only protein/gene information or the molecular structure of coronavirus that can be targeted by repurposable drugs. This means that unfortunately clinical information has not been utilized. Because COVID-19 patients with comorbidity disease have a high risk of illness, this unexplored relationship of comorbidity disease may help to find repurposable drugs.

In this study, we propose a network-based drug repurposing framework to predict candidate therapeutic agents for COVID-19. Our framework combines heterogeneous relational data such as disease-gene associations, disease-drug associations, drug-target gene associations, and comorbidity information with the novel coronavirus. Initially, we construct the comprehensive disease-gene-drug association networks by combining the disease-disease networks, gene-gene networks, and drug-drug networks. Then, the network propagation algorithm is applied to predict candidate repurposable drugs for COVID-19.

In more detail, from the associations between diseases, genes, and drugs, we construct a disease-gene-drug network (DGDr-Net) to understand the interactions between diseases, genes, and drugs. Then, the comorbidity information was added to the DGDr-Net to enhance the interactions between COVID-19 and other diseases. The constructed networks have tripartite relationships among diseases, genes, and drugs and these interactions can be represented in the form of layered networks. In establishing a tripartite relationship between a specific disease, gene, and drug, the following three procedures should be considered for drug repurposing: (1) Find a disease that is genetically or clinically similar to COVID-19 (at the disease level); (2) Find the key gene that serves as the bio-marker or some other therapeutic evidence for COVID-19 at the gene level; (3) Find a drug that can be used to treat COVID-19 by combining or finding a drug that is already being used to treat other diseases. Given the nature of tripartite relationships, all three steps need to be done simultaneously rather than independently. Thus, we construct a disease-gene-drug network consisted of three single-layered networks: a disease-disease network (disease layer), a gene-gene network (gene layer), and a drug-drug network (drug layer). Additionally, each single network is connected according to association information. Within the network, drug scores for COVID-19 are calculated by network-based label propagations that are using graph-based semi-supervised learning. Since the treatments of COVID-19 are not yet found, there is no label information associated with COVID-19 in the drug network. In those cases, the semi-supervised approach that can be applied when label information is insufficient is a very suitable method for recommending candidate drugs[14, 15]. Semi-supervised learning can deal with few labeled data and can perform predictions by propagating the label information to unlabeled nodes along with edges[16]. When the COVID-19 in the disease layer is set as a seed node for propagation, then the label information is transmitted through the edges of the disease-gene-drug network. The comorbidity and disease-gene association information in the disease layer are propagated to the gene and drug layer to inform the priority of the repurposable drug of COVID-19. From the resulting scores, the priorities of candidate drugs can be identified. The top-tier candidate drugs are recommended by repurposable drugs for coronavirus. The predicted results can help reduce trial-and-error in finding treatment for COVID-19.

## 2. Methods

We propose a network-based drug repositioning method that utilizes and integrates biochemical and clinical COVID-19 data such as disease comorbidity, protein associations, and drug-target gene information. The overall framework to find repurposable drugs for treating COVID-19 consists of mainly two parts as shown in Figure 1: (a) Constructing a Disease-Gene-Drug network and (b) Scoring for candidate repurposable drugs. We first construct the Disease-Gene-Drug networks (DGDr-Net) to figure out the interactions between the biological components related to COVID-19. The constructed networks have tripartite relationships among diseases, genes, and drugs and these interactions can be represented in the form of layered networks. Next, drug scores for COVID-19 are calculated by network-based label propagations using graph-based semi-supervised learning[14, 15, 17]. From the resulting scores, the priorities of candidate drugs can be identified. The top-tier candidate drugs are recommended by repurposable drugs for coronavirus.

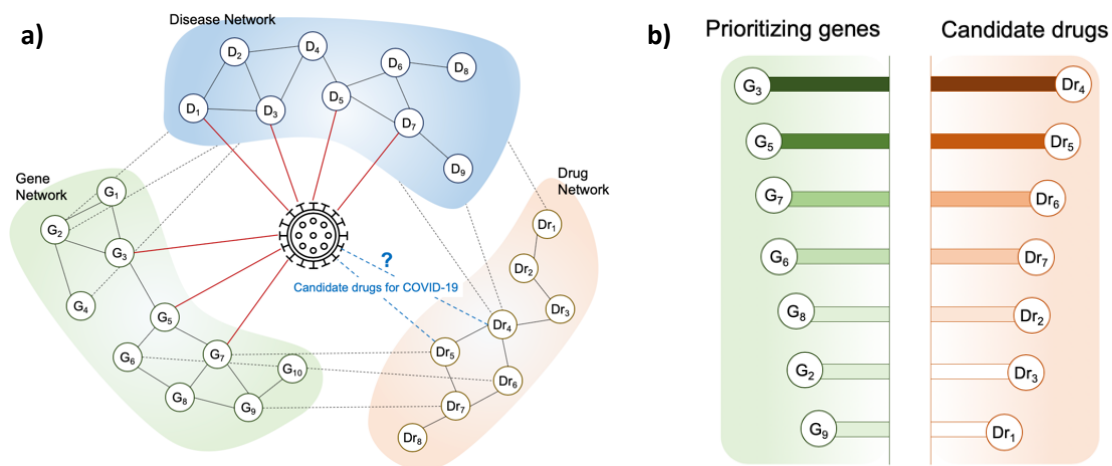

**Figure 1. Schematic description for the proposed method**: a) Disease-Gene-Drug networks with COVID-19, b) scoring results with genes/drugs for targeting COVID-19,

## 2.1 Disease-Gene-Drug network

We construct the Disease-Gene-Drug Networks (DGDr-Net) to show the COVID-19 centric interactome with components for diseases, genes, and drugs. The interactions among them have a tripartite relationship that can be represented in the form of layered networks. The DGDr-Net is a graph, $G = (V, E, S)$, where $V$ represents the set of nodes, $E$ represents the set of edges, and $S$ represents the set of layers. The proposed networks have three single layers with $S = \{S_D, S_G, S_{Dr}\}$, depending on the type of components (e.g., disease, gene, and drug) as shown in Figure 1(a). In the network, let $v_i \in V$ denote a node and $e_{ij} \in E$ to denote the edge that connects two vertices $v_i$ and $v_j$. The edge weights in the single layer have weighted values and unweighted values between different layers. The edge weights in a single layer $w_{ij}$ between two nodes $v_i$ and $v_j$ is calculated by Gaussian similarity kernel as follows:

$$w_{ij} = \begin{cases} \exp(-\frac{\text{dist}(v_i, v_j)}{\sigma^2}) &, v_i \sim v_j \\ 0 &, v_i \nsim v_j \end{cases} \tag{1}$$

where $\text{dist}(v_i, v_j)$ is a distance measure for adjacent nodes ($v_i \sim v_j$). The edge values between different layers can take a value 1 if the associations exist, otherwise 0. Then, the value of the similarity for DGDr-Net is represented by a matrix $\boldsymbol{W} = \{w_{ij}\}$. To easily describe the DGDr-Net with three single layers, the similarity matrix $\boldsymbol{W}$ can be expressed in block-wise as follows:

$$\boldsymbol{W} = \begin{bmatrix} W_D & W_{D\sim G} & W_{D\sim Dr} \\ W_{D\sim G}^T & W_G & W_{G\sim Dr} \\ W_{D\sim Dr}^T & W_{G\sim Dr}^T & W_{Dr} \end{bmatrix} \tag{2}$$

The block diagonal matrix ($W_D, W_G,$ and $W_{Dr}$) represents a similarity for single networks with disease (D), gene (G), and drug (Dr) layers respectively, and the block off-diagonal matrix represents the connections between different layers.

## 2.2 Scoring algorithms for repurposing

With the DGDr-Net, a scoring algorithm is applied to find repurposable drugs. Predicted resulting scores can be obtained simultaneously from each node of the disease layer, gene layer, and drug layer. In the disease layer, it is possible to obtain quantitatively how many associations between target and other diseases are. Disease scores for COVID-19 reflect both shared gene information with other diseases and comorbidity information. In the gene layer, disease-disease associations of the disease layer and protein interaction of the gene layer are propagated through the edges, and the resulting score indicates the priority of candidate target genes for treating COVID-19. In the drug layer, label information from COVID-19 in the disease layer is propagated through the edges of DGDr-Net. Resulting drug scores can identify the priority of candidate drugs for drug repurposing. Graph-based semi-supervised learning (SSL) is employed for scoring algorithms, as only one disease node (COVID-19) is given. SSL can perform predictions by propagating the label information to the nodes along with the edges even if there is only one labeled node[18].

The formulation of the scoring algorithm using SSL is as follows. Consider a DGDr-Net, $G = (V, E, S)$, with node $V(= V^D \cup V^G \cup V^{Dr})$ corresponding to the $n(= n_D + n_G + n_{Dr})$ vertices. Let $y = (y_1, \ldots, y_n)^T$ denote the label set of nodes, $f = [f^D, f^G, f^{Dr}]^T = (f_1^D, \ldots, f_{n_D}^D, f_1^G, \ldots, f_{n_G}^G, f_1^{Dr}, \ldots, f_{n_{Dr}}^{Dr})^T$ denote the set of resulting scores, and $W$ be a similarity matrix of DGDr-Net. Unlike the general classification problem where the target variable has a binary label ('+1' or '-1'), the problem setting of scoring in a semi-supervised approach has a unary label ('+1'). More specifically, a disease node of COVID-19 19 ($v_{COVID}^D$) is set to a unary label $y_{COVID} \in \{+1\}$, and the other nodes set to zero ($y \backslash y_{COVID} \in \{0\}$). There are two assumptions: a loss function (predicted scores in unlabeled nodes should be close to the given label of $y_i$ in labeled nodes) and smoothness condition (predicted scores in adjacent unlabeled nodes should not be too different). These assumptions are reflected in the value of $f$ by minimizing the following quadratic function with the graph Laplacian matrix $L$:

$$\min \ (f - y)^T(f - y) + \mu f^T L f \tag{3}$$

The graph Laplacian $L$ is defined as $L = D - W$, where $D = \text{diag}(d_i)$, and $d_i = \sum_j w_{ij}$. The user-specified parameter $\mu$ trades off loss and smoothness (to reduce complexity, $\mu$ is set to $1/\|L\|_1$). Then, the solution of this problem becomes $f = \left(I + \frac{L}{\|L\|_1}\right)^{-1} y$. The resulting score $f$ is rearranged in each layer: $f' = \frac{f - \min(f)}{\max(f) - \min(f)}$

To recommend the repurposable drugs, $n_{Dr}$ drugs are sorted in descending order according to the resulting score in drug layers. The rankings in drug lists can be analyzed in two ways: (1) to suggest possible candidate repurposable drugs and/or (2) to reduce the trial-and-error nature of current and future COVID-19 clinical trials.

### 3. Results of Network Construction
### 3.1 Data sources

We collected a list of components and relational data from public databases. Table 1 describes the data sources used to construct DGDr-Net. For a single disease network, 5,819 diseases and 31,991 disease-gene associations were obtained from the comparative toxicogenomics database (CTD)[19]. To screen diseases strongly associated with COVID-19 and to construct a more accurate COVID-19 centric disease network, community detection was applied[20]. 592 diseases were selected for disease nodes. To construct a gene network, we use the protein-protein interactions (PPIs) from the Search Tool for the Retrieval of Interacting Genes/Proteins (STRING) databases. The STRING DB collects and integrates every possible publicly available PPIs from gene neighborhood, gene fusion, gene co-occurrence, gene co-expression, experiments/biochemistry, and annotated pathway data[21]. To avoid false positive information, we select 26,681 genes (proteins) and 841,068 interactions with the high confidence level ($\geq 0.7$). For a drug network, 9,540 drugs (compounds) and 525,207 drug-target gene associations were obtained from DrugBank and CTD databases[19, 22].

COVID-19 is associated with 17 genes for biomarker or therapeutic evidence (e.g., CCL2, TNF, IL10, CXCL8, IL6, IL1B, AGT, IL2, CXCL10, CCL3, TMPRSS2, IL7, IL2RA, CSF3, TMPRSS4, ACE2, and BSG) as shown in Figure 2a. Also, 194 drugs in our network are currently in clinical trials for COVID-19[23]. A more detailed description of data and how to construct networks is explained below.

**Table 1. List of data sources to construct DGDr-Net.**

| | Description | # of data | Data sources | Released date |
|---|---|---|---|---|
| **Components** | Diseases | 592 out of 5,819 | CTD, MeSH descriptor | 2020-03-29 |
| | Genes (Proteins) | 26,681 | STRING | 2018-12-20 |
| | Drugs (Compounds) | 2,173 | DrugBank, CTD | 2020-04-22 |
| **Relational data** | Disease-Gene associations | 31,991 | CTD | |
| | Disease-Drug associations | 76,889 | DrugBank, CTD | |
| | Protein-Protein Interactions | 841,068 | STRING | |
| | Drug-Gene associations | 9,540 | DrugBank, CTD | |
| | Comorbidity diseases | 18 | 2-7 | |

**3.2 Disease network construction**

The disease network is a sub-graph $G = (V_D, W_D)$ when $S = \{S_D\}$ in the multi-layered DGDr-Net. A set of nodes $V_D$ represent a disease and $W_D$ represents the similarity between diseases. To obtain the COVID-19 centric disease network, we construct and integrate two types of networks; (a) networks with disease-gene associations and (b) networks with comorbidity relations. In the framework for drug repurposing using disease interactome, the former network allows us to use genetic information for other related diseases while the latter network allows us to utilize clinical data. First, we constructed networks with disease-gene associations using 5,819 diseases and 31,991 disease-gene associations. Each disease is represented as 26,681-dimensional gene vectors with binary attributes. If a disease is associated with a gene, its respective element in the disease vectors has a value ('1'), otherwise ('0'). (Note that attribute value 0 in disease vector does not mean that disease has no relationship with a certain gene. It could occur that the association has not been found yet). With 5,819 disease vectors, the edge weights for networks with disease-gene associations, $W_{D1}$, were calculated by Eq(1) using the cosine distance $\text{dist}(v_i, v_j) = 1 - (\boldsymbol{v}_i \cdot \boldsymbol{v}_j)/(\| \boldsymbol{v}_i \| \cdot \| \boldsymbol{v}_j \|)$. Next, we constructed a network with comorbidity relations involving the 18 aforementioned diseases highly comorbid with COVID-19. The 18 aforementioned comorbid diseases were curated by recent researches[2-7]. The edge weights for comorbidity network, $W_{D2}$, has binary values, each of which stands for reported comorbidity relation ('1') or not ('0'). Figure 2(b) depicts a comorbidity network with 18 comorbidity diseases and COVID-19. To screen tightly associated diseases with COVID-19 and construct more eligible disease networks, community detection was applied[20]. Two modules include COVID-19 and 18 comorbidity diseases. Finally, 592 diseases in two modules were used in the disease network. With two types of networks, network integration was performed. The combined similarity $W_D$ is represented as a linear combination of both networks with coefficient $\beta_1$ and $\beta_2$. To find the coefficients $\boldsymbol{\beta}$ for the linear combinations, grid search was performed in the range (0.1, 1) based on $\beta_1$. While adjusting the coefficients, DGDr-Net was applied a scoring algorithm to aim for recommending clinical trial drugs for COVID-19, and average precision was used as a metric. As a result, we choose $\beta_1 = 0.3$ for networks with disease-gene associations and $\beta_2 = 0.7$ for networks with comorbidity relation. Final disease networks are shown in Figure 2(c).

### 3.3 Gene network construction

Gene network is a sub-graph, $G = (V_G, W_G)$, when $S = \{S_G\}$ in DGDr-Net. The protein-protein interaction (PPIs) is employed for gene networks from the STRING database[21]. To avoid false positive information, we select 26,681 genes (proteins) and 841,068 interactions with a high confidence level ($\geq 0.7$). In the gene network, the edge stands for protein interaction, in which the edge weight indicates the presence or absence ('1' or '0' respectively).

## 3.4 Drug network construction

Drug network is a sub-graph, $G = (V_{\text{Dr}}, W_{\text{Dr}})$, when $S = \{S_{\text{Dr}}\}$ in multi-layered DGDr-Net. 9,540 drugs (compounds) and 525,207 drug-target gene associations were obtained from DrugBank and CTD databases[19, 22]. Each drug is composed of 26,681-dimensional binary attributes, each of which stands for existing target gene associations ('1') or not ('0'). The similarity matrices for the drug network $W_{\text{Dr}}$ were calculated by cosine distance and transformed by Gaussian similarity kernel by Eq(1).

## 3.5 Connections between different single layers

Recall similarity matrix $W$ for DGDr-Net in Equation (2). The block off-diagonal matrix represents the connections between different layers. From the relational data in Table 1, the connection weights between different single layers have binary values of 1 if there are associations of two nodes in different layers, 0 otherwise.

## 4. Scoring results for repurposable drugs and clinical implication

## 4.1 Experimental Settings

We performed a scoring algorithm to predict candidate repurposable drugs for COVID-19 using tripartite relationships of disease, gene, and drug in DGDr-Net. It can align drugs through the connections between nodes in the network with label propagations. Currently, no therapeutic drugs for coronavirus have been developed, which means that there are no ground truths that can be used as labeled data in machine learning perspective. SSL can deal with few labeled data and perform prediction by propagating the label information to the nodes in three layers along with edges[18]. To find repurposable drugs, we performed the scoring algorithms with a label for COVID-19 only in the disease layer. Also, all of the drug information related to COVID-19 were assumed to be unknown which means the edge connections between COVID-19 and all of the drugs were not included in the DGDr-Net. Out of 2,173 highlighted drugs, 194 drugs are currently in clinical trials for COVID-19[23]. To verify the performance of our scoring results, we assume these 194 drugs as ground truths. Only one node (COVID-19) in the disease layer was set as "1" for the labeled set, and the remaining 590 diseases, 26,681 genes and 2,173 drugs were set as "0" for the unlabeled set. After scoring, we tested how many drugs in clinical trials were highly ranked.

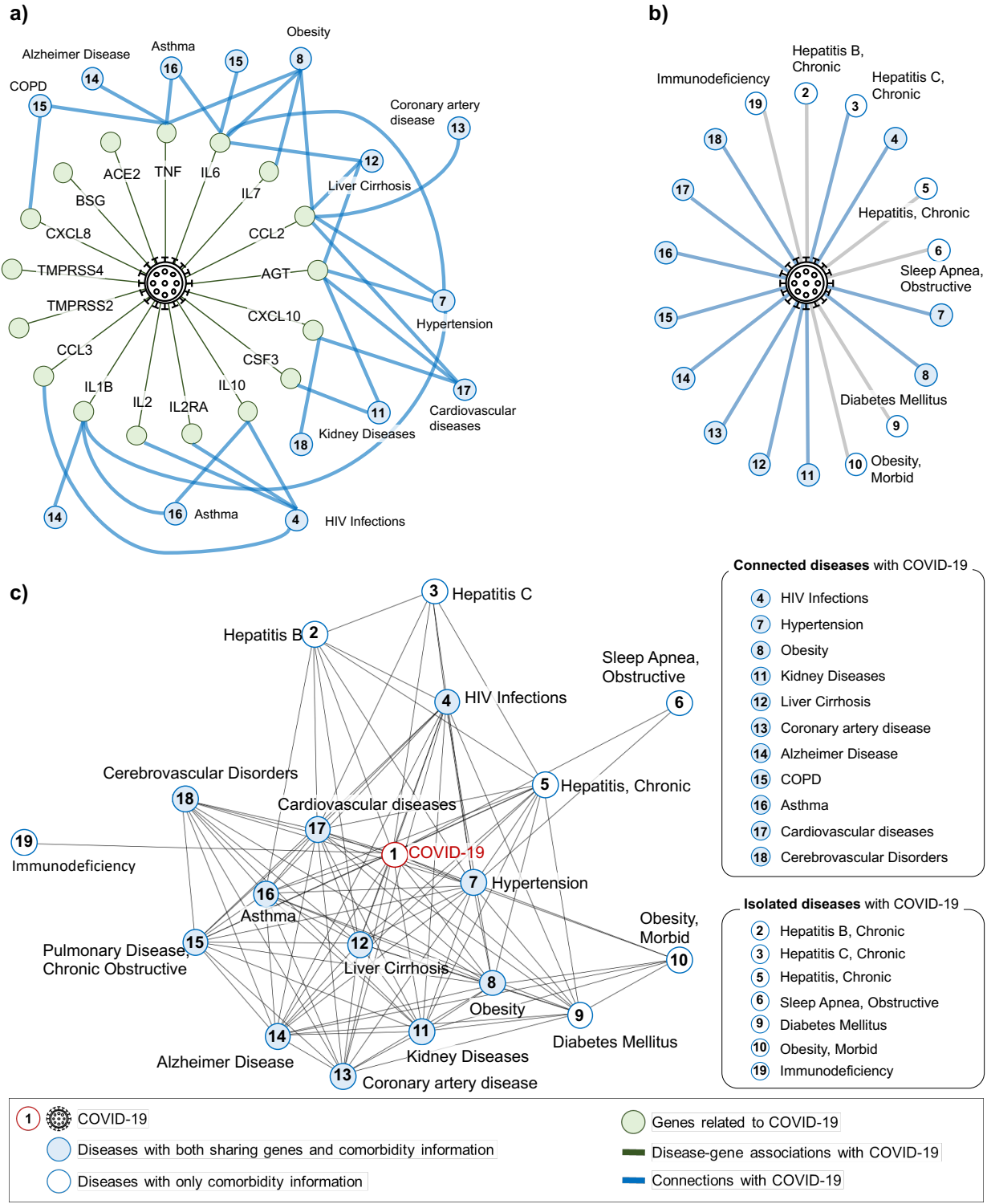

Figure 2. Snapshot of the disease network focused on COVID-19 and 18 comorbidity diseases: a) Disease-gene network, b) Reported comorbidity relationships with COVID-19, c) combined disease network

## 4.2 Implication and Validation of DGDr-Net

The scoring algorithm with DGDr-Net provides score values for all nodes (disease, gene, drugs) in the three layers. Disease scores represent which diseases are strongly associated with COVID-19, and gene scores represent the priorities of target genes for biomarkers. Also, drug scores represent the priorities of candidate drugs for COVID-19. Figure 3 shows the list of top 30 ranked diseases, genes, and drugs when COVID-19 is given. The colored nodes represent COVID-19 related diseases, genes, and drugs, and they are sorted in descending order according to the resulting scores in each layer. Blue nodes indicate diseases, green nodes indicate the genes, and red nodes indicate the drugs that are recommended for repositioning. Also, we show the tripartite connections among diseases, genes, and drugs such as the disease-gene associations and drug-disease associations.

Figure 3(a) represents the associations between 30 high ranked diseases (blue nodes) and genes (green nodes) focused on COVID-19 (black node). The blue-colored circles indicate diseases with comorbidity information, and the blue outlined circles indicate diseases connected to COVID-19 from sharing gene information. 18 diseases with comorbidity information are at the top of the ranking. Among them, 10 diseases (asthma, chronic obstructive pulmonary disease (COPD), HIV infections, cardiovascular diseases, cerebrovascular disorders, hypertension, hepatic cirrhosis, obesity, Alzheimer's disease, and coronary artery disease) have strong associations with COVID-19 because they not only have common genes with COVID-19 but also have additional comorbidity information factored in to the analysis. 8 diseases (kidney diseases, morbid obesity, common variable immunodeficiency, obstructive sleep apnea, chronic hepatitis B, chronic hepatitis C, chronic hepatitis, and diabetes mellitus) were connected with COVID-19 primarily via the comorbidity information. Among the comorbid diseases reported as an underlying medical condition, there was generally a positive correlation between the number of genes shared with COVID-19 and the given score. The remaining 12 diseases are linked by sharing gene hypothesis, and diseases with high edge weights have a high score. The score results in the gene layer is largely interpreted in two ways: (1) the priorities of the strongly associated genes with COVID-19 and (2) a list of genes that can be candidate target genes for treatments. The priority of already known genes with COVID-19 is the result of the label propagation by reflecting both the disease interactions and protein interactions.

Figure 3(b) represents the associations between diseases and candidate drugs. The red-colored circles in the drug lists denotes currently in clinical trial drugs and red outlined circles are currently not yet. As mentioned above, we have not yet developed a COVID-19 therapeutic agent. Thus, we recommend all 30 highlighted drugs as candidate repositioning drugs. The rankings in drug lists can be analyzed in two ways: (1) The results can suggest the possible candidate repurposable drugs from the rank, (2) To reduce the trial-and-error of clinical trials, the list of drugs can be compared with current clinical trial drugs for COVID-19.

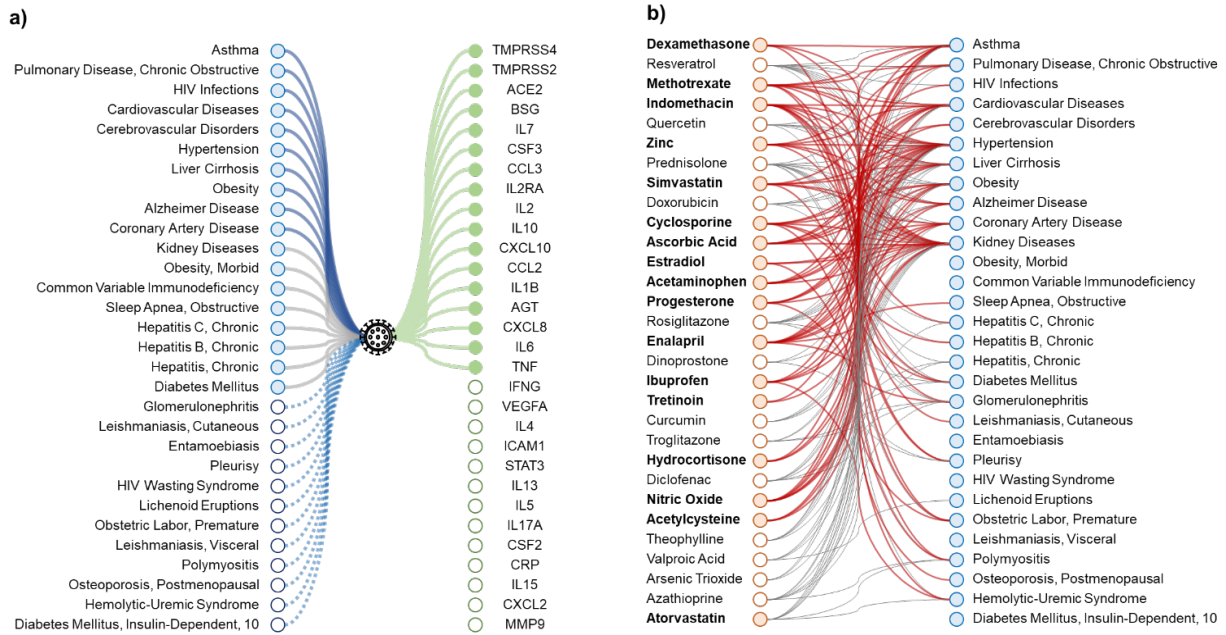

**Figure 3. Associations between top 30 ranked diseases, genes, and drugs**

### 4.3 Clinical utility of the proposed drugs

Figure 4 depicts the scoring results of top candidate drugs for COVID-19. On the graph, the solid gray line stands for the value of scores $f$ of the candidate drugs in descending orders. Out of 2,173 drugs, we assume that these drugs should be repurposable drugs for COVID-19. The red circles on the line correspond to the current clinical trial drugs, and the black one is not. Steroids such as dexamethasone, prednisolone, and hydrocortisone were recommended as top candidates. Among them, dexamethasone, an anti-inflammatory drug, had the highest scores. Dexamethasone is a very cheap steroid that reduces inflammation by mimicking anti-inflammatory hormones produced by the body. COVID-19 treatment guidelines recommend using dexamethasone 6 mg per day for up to 10 days for hospitalized patients with COVID-19[24, 25]. On the score curve, the inserted drugs between clinical-trial drugs are regarded as strong candidates even though they are not used in clinical trials.

Several proposed drugs need to be considered for clinical application. First, dexamethasone, prednisolone, and hydrocortisone are steroid analogues, which affect immune and inflammatory functions. There have been reports suggesting that steroids may be effective in the control of systemic inflammation or 'cytokine storm' in severe COVID-19 cases[25], although there are several on-going trials on the effectiveness of steroid treatment[10]. The current study also supports the possibility of steroid therapy in patients with COVID-19. Until now, there has been controversy regarding the use of NSAIDs in COVID-19 patients. Although several literatures warned the use of NSAIDs, others disagreed[26]. With acetaminophen being included as one of the highlighted drugs in our study, we wish to bring attention to the role NSAIDs may play in helping an individual with COVID-19. Given the body's inflammatory response to the virus, researchers have been studying the effects of some immune-modulating drugs including methotrexate, cyclosporin[27, 28], although there is a paucity of information on other immune-modulating medications or cytotoxic drugs including azathioprine, doxorubicin, valproic acid, and Arsenic Trioxide.

Based on the current study, further studies are needed to evaluate the possibility of immune-modulating drugs in the context of COVID-19. Several proposed drugs also need attention such as ACE inhibitors (enalapril), lipid lowering medications (simvastatin), hormonal medication (estradiol, progesterone) and antidiabetic drugs (rosiglitazone, troglitazone). These medications may be more effective in specific populations with specific comorbidities such as kidney disease, diabetes, or coronary/cardiovascular disease. We may also have to evaluate the efficacy of these medications in specific populations. However, of course, before any of these potential treatments are given to help patients suffering from COVID-19, rigorous clinical trials are required.

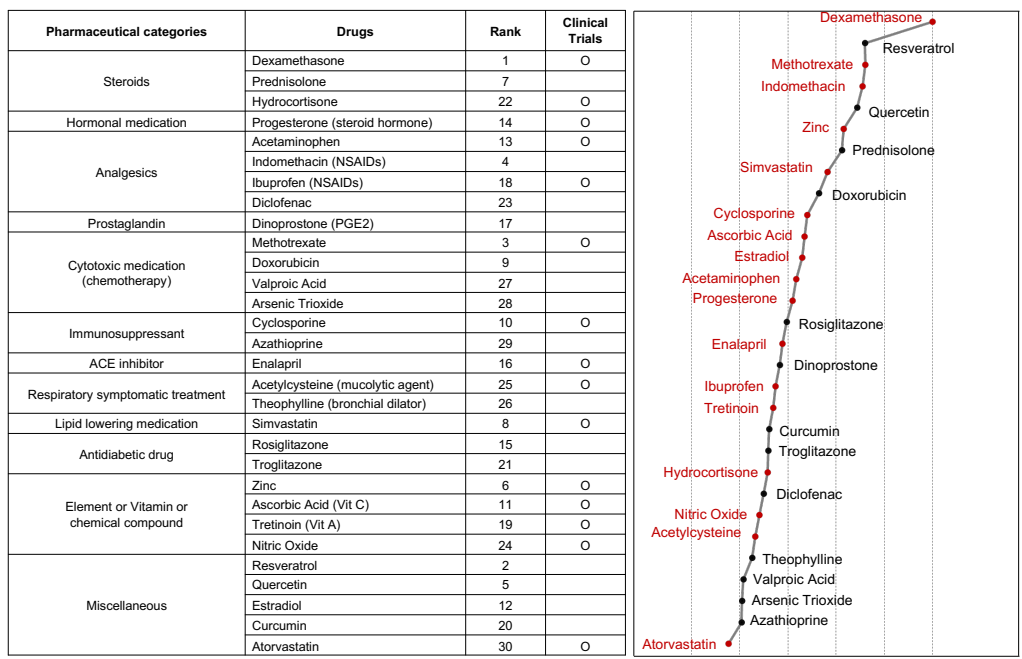

| Pharmaceutical categories | Drugs | Rank | Clinical Trials |
|---|---|---|---|
| Steroids | Dexamethasone | 1 | O |
| | Prednisolone | 7 | |
| | Hydrocortisone | 22 | O |
| Hormonal medication | Progesterone (steroid hormone) | 14 | O |
| Analgesics | Acetaminophen | 13 | O |
| | Indomethacin (NSAIDs) | 4 | |
| | Ibuprofen (NSAIDs) | 18 | O |
| | Diclofenac | 23 | |
| Prostaglandin | Dinoprostone (PGE2) | 17 | |
| Cytotoxic medication (chemotherapy) | Methotrexate | 3 | O |
| | Doxorubicin | 9 | |
| | Valproic Acid | 27 | |
| | Arsenic Trioxide | 28 | |
| Immunosuppressant | Cyclosporine | 10 | O |
| | Azathioprine | 29 | |
| ACE inhibitor | Enalapril | 16 | O |
| Respiratory symptomatic treatment | Acetylcysteine (mucolytic agent) | 25 | O |
| | Theophylline (bronchial dilator) | 26 | |
| Lipid lowering medication | Simvastatin | 8 | O |
| Antidiabetic drug | Rosiglitazone | 15 | |
| | Troglitazone | 21 | |
| Element or Vitamin or chemical compound | Zinc | 6 | O |
| | Ascorbic Acid (Vit C) | 11 | O |
| | Tretinoin (Vit A) | 19 | O |
| | Nitric Oxide | 24 | O |
| Miscellaneous | Resveratrol | 2 | |
| | Quercetin | 5 | |
| | Estradiol | 12 | |
| | Curcumin | 20 | |
| | Atorvastatin | 30 | O |

**Figure 4. Score curve of top 30 repurposable drugs for COVID-19**

## 5. Conclusions

In this study, we developed a network-based drug repurposing framework for recommending repurposable drugs by utilizing and combining COVID-19 related data such as disease comorbidity, protein associations, and drug-target gene information. We first constructed a form of layered disease-gene-drug networks (DGDr-Net) to figure out the interactions among the diseases, genes, and drugs related to COVID-19. With the network, we apply semi-supervised scoring algorithms to identify candidate repurposable drugs. However, the information we have about coronavirus associated drugs is not yet sufficient because we do not yet have a cure for COVID-19. Therefore, in the scoring procedures, similar to the Cold Start Problem in recommendation systems, the candidate drugs were identified without any inferences between COVID-19 and drugs.

We provided the priorities of candidate repurposable drugs and candidate target genes based on predicted scores. Of the top 30 drugs, 17 drugs were currently in clinical trials for the treatment of COVID-19. Steroids such as dexamethasone, prednisolone, and hydrocortisone are recommended as top candidates. Among them, dexamethasone, an anti-inflammatory drug, had the highest scores.

A unique strength of our experimental approach is the inclusion of comorbidity factors to our network. This allowed us to more accurately identify repurposable drugs while affording a bird's-eye view of the intricate associations between COVID-19 and other related diseases, drugs, and genes. One particular limitation of our study is the relatively small sample size of the few databases we utilized; however, this concern is quickly alleviated as the robust yet flexible nature of a network-based approach allows us to very easily supplement and correct our current model. As we receive the newest information regarding the novel coronavirus, we can easily update the candidate drug/gene components of our networks, performing a set of updated calculations and generating an updated gene and drug candidate list almost instantly. With this in mind, we hope our approach may help clinicians and scientists make the hard decisions regarding which drugs or gene targets to test first in this global race for a cure.